\documentclass[a4paper]{article}

\usepackage{icrc2013}
\usepackage[english]{babel}

\title{Are most of the VHE gamma-ray unidentified sources relic PWNe?}

\shorttitle{unids-PWNe?}

\authors{
Omar Tibolla$^{1}$,
Michael Vorster$^{2}$,
Sarah Kaufmann$^{3}$,
Stefan Ferreira$^{2}$,
Karl Mannheim$^{1}$.
}

\afiliations{
$^1$ ITPA, Universit\"at W\"urzburg, Campus Hubland Nord, Emil-Fischer-Str. 31 D-97074 W\"urzburg, Germany \\
$^2$ Centre for Space Research, North-West University, Potchefstroom Campus, 2520 Potchefstroom, South Africa \\
$^3$ Landessternwarte, Universit\"at Heidelberg, K\"onigstuhl, D 69117 Heidelberg, Germany 
}

\email{omar.tibolla@gmail.com ; Omar.Tibolla@astro.uni-wuerzburg.de}

\abstract{The evolution of Pulsar Wind Nebulae (PWNe) plays a crucial role in interpreting the very high energy (VHE; $> 10^{11}$ eV) gamma-ray unidentified sources; and moreover it represents the only viable option to explain the discovery of several ``dark sources'' in the TeV gamma-ray (i.e. VHE gamma-ray sources without lower energies, radio or X-ray counterparts).

The newest time-dependent modeling of PWNe presented in $[1]$ and $[2]$ has to be tested on a broader sample of young well-known PWNe and applied to the full-sample of ``dark sources''.

The consequences of this interpretation go far beyond the interpretation of ``dark sources'': e.g. there could be strong implication in the origin of cosmic rays and (when considering a leptonic origin of the gamma-ray signal) they can be important for reinterpreting the detection of starburst galaxies in the TeV gamma-ray band.}

\keywords{Unidentified gamma-ray sources, Pulsar Wind Nebulae, HESS J1507-622, HESS J1837-069, HESS J1616-508, HESS J1702-420, HESS J1708-410, HESS J1804-216.}

\begin{document}
\maketitle

\section{Unidentified sources as Pulsar Wind Nebulae}

Since decades Pulsar wind nebulae (PWNe) are studied as unique laboratories for the highly efficient acceleration of particles to ultra-relativistic energies: e.g. in the late sixties ideas about particle acceleration by electromagnetic waves emitted by pulsars $[3]$ led to the discovery that a significant fraction of the spin-down power of the Crab pulsar is dissipated through a relativistic wind with in situ acceleration of particles at its termination shock, since the cooling time scales of the synchrotron-emitting electrons are very short in X-rays $[4]$.

40-50 years after their discovery great progress has been made in understanding PWNe (e.g. the evidence for the accumulation of very-high energy electrons in a PWN in many TeV PWNe, such as HESS J1825-137 $[5]$, is crucial for the points treated here), but many questions remain.
In particular, since several years and especially after the discovery of the off-set VHE unidentified source HESS J1507-622 $[6]$ $[7]$, it was suggested that an evolved PWN can indeed lead to a fairly bright gamma-ray source without any lower energy (especially X-ray) counterpart (e.g. $[7]$, $[8]$, $[9]$).

This scenario (so-called ``ancient PWN scenario'') was supported by MHD simulations of composite SNRs $[10]$ and found evidence in the observations (e.g. $[5]$): in particular the discovery of HESS J1507-622 played a crucial role, since an ancient PWN is the only viable option to describe the VHE emission of this source so far (e.g. $[7]$, $[9]$, $[11]$).

\section{Time-dependent modeling of PWNe and the complete sample}

Recently, we developed a new spatially-independent model to calculate the temporal evolution of the electron/positron spectrum in a spherically expanding PWN $[1]$ $[2]$.
This new code has been tested $[1]$ $[2]$  on the young PWN G21.5-0.9, and successfully applied to the two unidentified VHE gamma-ray sources HESS J1427-608 and HESS J1507-622, strengthening the argument that the unidentified VHE gamma-ray sources can indeed be identified as aged PWNe.

In $[2]$ it is mentioned the necessity to test the model to a larger number of known PWNe in order to derive a statistically significant set of parameters that can be used as a guideline for PWN evolution; moreover, this additional test allows us to constrain better the unidentified VHE gamma-ray sources in the frame of PWN evolution and, eventually, to discard the unidentified sources that do not fit in this picture. Some preliminary results of this on-going work are presented in the next section.

The first step consisted in having a complete sample of the TeV bright PWNe, the candidate PWNe and the unidentified VHE gamma-ray sources; the complete sample is shown in Table 1. The ``border line'' between candidate PWNe and unidentified sources is obviously rather faint (i.e. the candidates are by definition not firmly identified): therefore here we consider as candidate PWNe only the sources that have strong indications of their PWN nature and no other likely scenario to explain their VHE gamma-ray emission.

\begin{table*}[h]
\begin{center}
\begin{tabular}{|c|c|}
\hline PWNe (and PWN candidates) & unidentified VHE gamma-ray sources \\ \hline
CTA 1  & HESS J1018-589 \\                     
Crab   & HESS J1427-608 \\ 
N 157B   & HESS J1457-593 \\ 
Geminga & HESS J1503-582 \\
G21.5-0.9 & HESS J1507-622 \\
Vela X & HESS J1626-490 \\
HESS J1026-582 & HESS J1614-518 \\
HESS J1119-614 &  HESS J1616-508  \\
HESS J1303-631$^1$ &  HESS J1641-463\\
HESS J1356-645 & HESS J1632-478 \\
Rabbit nebula & HESS J1634-472 \\
Kookaburra & HESS J1646-458 (Wd1?) \\
HESS J1458-608 & HESS J1702-420$^2$ \\
MSH 15-52 & HESS J1708-410 \\
G327.1-1.1 & HESS J1729-345 \\
HESS J1640-465 & HESS J1741-302 \\
PSR B1706-44 & HESS J1745-303 (hot spot C)$^3$ \\
HESS J1718-385 & HESS J1747-248 (Terzan 5?) \\
HESS J1745-303 (hot spot B)$^3$ & HESS J1804-216 \\
HESS J1747-281 & HESS J1808-204 \\
HESS J1809-193 & HESS J1818-154 \\
HESS J1813-178 & HESS J1832-093 \\
HESS J1825-137 & HESS J1834-087 \\
HESS J1831-098 & HESS J1841-055 \\
HESS J1833-105 & VER J2016+372 \\
HESS J1837-069 & HESS J1843-033 \\
Kes75 & HESS J1848-018 \\
HESS J1857+026 & HESS J1849-000 \\
MGRO J1908+06 & HESS J1858+020 \\
HESS J1912+101 & G54.1+0.3 \\
Boomerang & HESS J1943+213 \\
~~~ & MGRO J2019+37 \\
~~~ & VER J2019+407 \\
~~~ & TeV J2032+4130 \\ \hline
\end{tabular}
\caption{The complete sample of VHE gamma-ray PWNe and unidentified sources that we are using in order to test our model. The sources are ordered following their R.A. . $^1$: HESS J1303-631 is the first H.E.S.S. unidentified source and recently $[12]$ it has been identified as an aged PWN. $^2$: HESS J1702-420 $[13]$ was initially thought to be a middle-aged PWN, however, after deep MWL observations, this scenario seemed unlikely. $^3$: for more details about this morphologically complex source $[14]$.}
\label{table_single}
\end{center}
\end{table*}

Moreover, it is interesting to note that some of the candidate PWNe shown in Table 1, as well as some unidentified sources, are also coincident with SNR shells and/or with molecular clouds: in these cases, the application of our model is even more important in order to constrain the source identification.

\section{Preliminary results}

Among the $\sim$60 VHE gamma-ray sources of Table 1, in this section we focus our attention on five bright VHE gamma-ray sources of the H.E.S.S. Galactic plane survey (GPS) $[15]$: 

\begin{itemize}
\item HESS J1837-069: since the H.E.S.S. GPS $[15]$ this source was considered a candidate PWN, since pulsed emission was detected from AX J1838.0-0655 $[16]$, suggesting a 70 ms pulsar as the origin of the nebula. {\it ASCA} observation was recently confirmed by {\it Suzaku} $[17]$, strengthening the idea that HESS J1837-069 is indeed a PWN. Our time-dependent PWN model can reproduce the Spectral Energy Distribution (SED) of this source (Fig. 1).
\item HESS J1616-508 $[15]$: it is often cited among the ``dark sources'', together with HESS J1427-608 and HESS J1507-622 (e.g. $[7]$, $[9]$), given the lack of lower energy counterparts. The similarities of HESS J1616-508 with the other two ancient PWNe are underlined by Fig.2.
\item HESS J1702-420 $[13]$ $[15]$ is a very peculiar source: while unidentified VHE gamma-ray sources are often identified as PWN, HESS J1702-420's history instead went in the opposite direction. In fact, its peculiar morphology and in particular its (significant)  emission ''tail`` pointing to the powerful (i.e. it would provide enough spin-down energy loss to produce the observed VHE gamma-ray emission) pulsar PSR J1702-4128 (e.g. $[13]$) suggested that this source could be an intermediate age PWN. However, deep X-ray observations seemed to disprove this thesis $[21]$ and suggested an hadronic origin for HESS J1702-420. Fig. 3 suggests that an aged PWN could explain the VHE gamma-ray emission from HESS J1702-420 as well.
\item HESS J1708-410 $[13]$ $[15]$: as well as HESS J1616-508, it is often cited among the ``dark sources'' (e.g. $[7]$, $[9]$), given the lack of X-ray and radio counterparts. Its similarities with the other ancient PWNe are underlined by Fig.4.
\item HESS J1804-216; for this source two alternative associations were proposed: with SNR G8.7-0.1 (or W30) or with the high-spin down pulsar PSR J1803-2137 (i.e. a slightly off-set PWN) $[15]$. The problems about this source are quite different from the ones mentioned here so far (i.e. lack of counterpart); in fact, the region around HESS J1804-216 shows several sources: the SNR W30, the compact remnant G8.31-0.09, and a number of additional structures identified as H II regions (e.g. $[23]$); its identification is therefore particularly unclear. Moreover, the recent detection of G8.7-0.1 by {\it Fermi}-LAT $[24]$ raised further questions: in fact, the GeV morphology and the GeV gamma-ray spectrum are not fitting well with the TeV gamma-rays; $[24]$ suggested the TeV gamma-ray signal originates from the interaction of particles accelerated in G8.7-0.1 with molecular clouds. Our model, as shown in Fig. 5, faces several problems in reproducing the SED of HESS J1804-216, suggesting a not likely PWN interpretation.
\end{itemize}

It is important to underline that the H.E.S.S. spectra are shown in Fig.1-5  with the statistical errors only; the systematic error on flux scale are estimated to 30\%: see $[15]$ for more details.

\begin{figure}[t]
  \centering
  \includegraphics[width=0.45\textwidth]{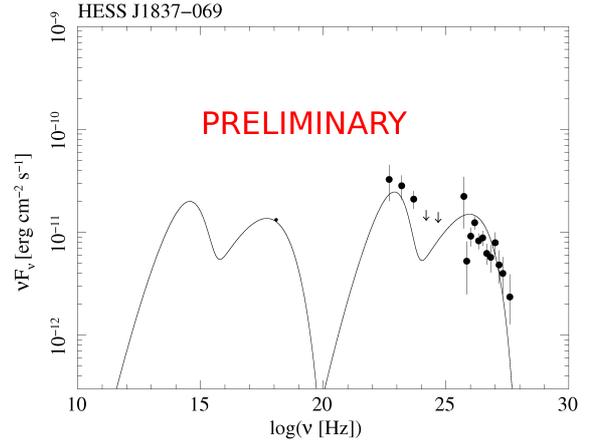}
  \caption{Model prediction for HESS J1837-069. H.E.S.S. data points are taken from $[15]$, {\it Fermi}-LAT data from $[18]$, {\it Suzaku} X-ray data from $[17]$; at radio wavelengths, the no source spatially compatible with HESS J1837-069 in the 2$^{nd}$ Epoch Molonglo Galactic Plane Survey (MGPS-2) $[19]$.}
  \label{1837}
 \end{figure}

\begin{figure}[t]
  \centering
  \includegraphics[width=0.45\textwidth]{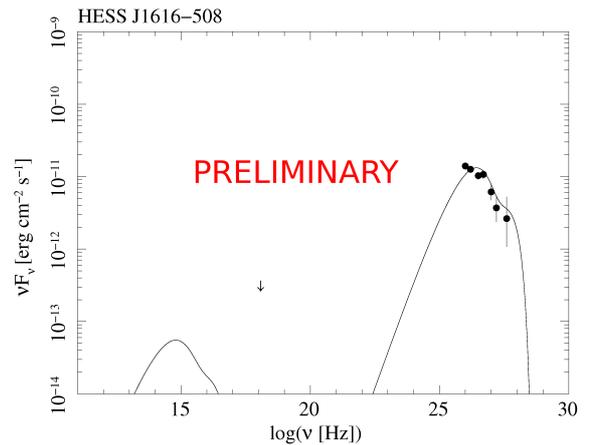}
  \caption{Model prediction for HESS J1616-508. H.E.S.S. data points are taken from $[15]$ and the X-ray {\it Suzaku} upper limit from $[20]$. }
  \label{1616}
 \end{figure} 

\begin{figure}[t]
  \centering
  \includegraphics[width=0.45\textwidth]{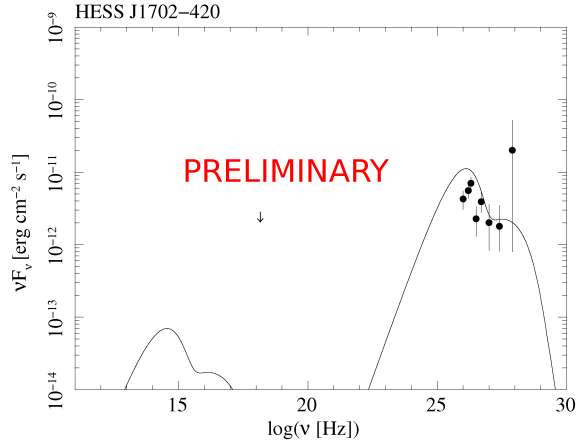}
  \caption{Model prediction for HESS J1702-420. H.E.S.S. data points are taken from $[15]$ and the 2-10 keV {\it Suzaku} upper limit ($F < 2.4 \times 10^{-12}$ erg cm$^{-2}$s$^{-1}$) from $[21]$.}
  \label{place}
 \end{figure}

\begin{figure}[t]
  \centering
  \includegraphics[width=0.45\textwidth]{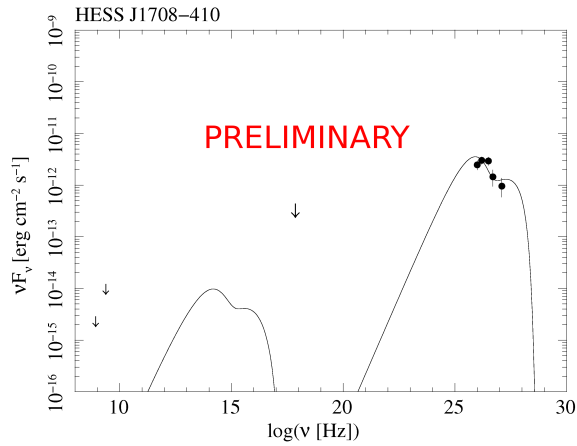}
  \caption{Model prediction for HESS J1708-410.  H.E.S.S. data points are taken from $[15]$; the radio (in the 843 MHz and 2.4 GHz bands) and X-ray ({\it XMM-Newton}; in the 2-4 keV band $ F < 3.2 \times 10^{-13}$ erg cm$^{-2}$s$^{-1}$) from $[22]$.}
  \label{place}
 \end{figure}

\begin{figure}[t]
  \centering
  \includegraphics[width=0.45\textwidth]{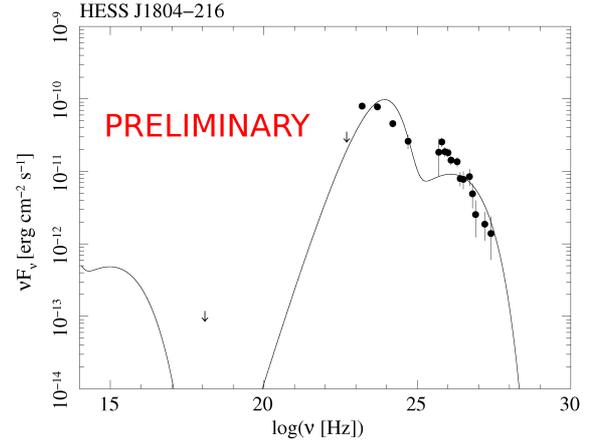}
  \caption{Model prediction for HESS J1804-216. H.E.S.S. data points are taken from $[15]$ and {\it Fermi}-LAT data from $[18]$. The X-ray upper limit is chosen as the lowest flux measured: $F(0.3-10 \mathrm{kev}) = 10^{-13}$ erg cm$^{-2}$s$^{-1}$) $[25]$.}
  \label{place}
 \end{figure}

\section{Conclusion and corollaries}

Even if the results shown here are still preliminary, the new time-dependent model (described in $[1]$ and $[2]$) is clearly a powerful tool to study PWNe, but not only; in fact it can help in strengthening the idea that most of the unidentified VHE gamma-ray sources can be described within a PWN framework and this will double (as underlined in Table 1) the population of VHE PWNe. In particular it strengthen the idea suggested by $[7]$ and $[8]$ that the so-called ``dark sources'' are indeed ancient PWNe.

Moreover strengthening the idea about the very long lifetime of inverse Compton emitting electrons in VHE gamma-ray PWNe could have strong implications, far beyond the interpretation of ``dark sources''.
It could, for example, have implications also about the origin of cosmic rays: in fact, in order to close possible gaps (such as the apparent small efficiency in accelerating cosmic rays, e.g. Cas A $[26]$ or Tycho $[27]$), people are currently searching for other possibilities. Moreover, looking the VHE sky, the dominant population of identified sources is not represented by shell-type SNRs, but PWNe are the most numerous. And indeed it was proposed that at the termination shock of the pulsar wind, also hadrons could be accelerated as well as leptons (e.g. $[28]$ $[29]$ $[30]$ $[31]$).

Moreover it has been underlined by $[32]$ that PWNe can be important for reinterpreting the detection of starburst galaxies in the TeV gamma-ray band.



\vspace*{0.5cm}
\footnotesize{{\bf Acknowledgment:}{The authors would like to acknowledge and honor the memory of a great scientist, a wonderful person: O.C. de Jager; moreover, the idea of applying the ancient PWN model ($[8]$) to a complete sample of unidentified VHE gamma-ray sources and some preliminary work started in collaboration with Prof. de Jager himself, while finalizing $[7]$ and $[8]$ .}}

\end{document}